\documentclass[10pt,b5paper,twoside]{padeu}   
\usepackage{epsfig,natbib_padeu}              

\begin{document}

\title{Canonical theory of the Kantowski-Sachs cosmological models}
\author{Zsolt Horv\'ath\inst{1}, 
Zolt\'{a}n Kov\'acs\inst{2}}
\authorrunning{Zs. Horv\'ath, Z. Kov\'acs}
\institute{\inst{1} Department of Theoretical Physics, University of Szeged, 6720 Szeged, D\'{o}m t\'{e}r 9, Hungary\\
	   \inst{2} Max-Planck-Institut f\"ur Astronomie, K\"onigstuhl 17. D-69117 Heidelberg, Germany}
\email{\inst{1}zshorvath@titan.physx.u-szeged.hu, 
       \inst{2}kovacs@mpia-hd.mpg.de}

\abstract{
We briefly discuss the Hamiltonian formalism of Kantowski-Sachs space-times with vacuum, anisotropic fluid and two cross-streaming radiation field sources. For these models a cosmological time is introduced. New constraints are found in which the fluid momenta are separated from the rest of the variables. In consequence their Poisson brackets give an Abelian algebra.
}

\maketitle

\section{Introduction}

The Kantowski-Sachs (hereafter KS) cosmologies have two symmetry properties,
the spherical symmetry and the invariance under spatial translations. The
vacuum KS solution is equivalent to the inner Schwarzschild space-time and
exact solutions were also found in presence of some matter fields for
homogenous cosmological models. Kantowski and Sachs [\citet{kansac}]
provided solutions for dust space-times but later on KS geometries with
other matter sources were found, such as scalar fields [\citet{barr}],
prefect fluid [\citet{perf}], anisotropic fluid [\citet{homogen}] and exotic
fluid [\citet{wormhole}] models. Here we give a short overview of the
Hamiltonian theory for KS cosmology in the case of vacuum and anisotropic
fluid sources. We employ the equivalence of the latter with the model
consisting of two, in- and outgoing radiation streams in a stellar
atmosphere. We introduce a cosmological time for the space-time containing
colliding radiation streams and introduce new constraints. These results
might represent an important step in carrying out a consistent canonical
quantization and in building up KS quantum cosmologies.

\section{Vacuum Kantowski-Sachs space-times}

The line element of KS space-times is given by 
\begin{equation}
ds^{2}=-d\tau^{2}+H\left( \tau\right) dh^{2}+R^{2}\left( \tau\right)
d\Omega^{2}\;\;,  \label{ds2}
\end{equation}
where $\tau$ is the cosmological time, $h$ a radial coordinate and $
d\Omega^{2}$ is the metric on the unit sphere. For vacuum this metric can be
written as 
\begin{equation}
ds^{2}=-d\eta^{2}+b^{2}\tan^{2}\eta dh^{2}+R^{2}d\Omega^{2}\;\;,
\label{dsvac}
\end{equation}
where we introduced the angle parameter $\eta$, 
\begin{equation}
\tau=a\left( \eta+\sin\eta\cos\eta\right) +c\;\;\qquad a,b,c\in I\!\!R\;,
\end{equation}
usually employed in homogenous, spherically symmetric cosmologies. With the
coordinate transformation $R(\eta)=a\cos^{2}\eta$ the solution (\ref{dsvac})
can be cast into the form of the Schwarzschild metric 
\begin{equation}
ds^{2}=-F(R)dT^{2}+F^{-1}(R)dR^{2}+R^{2}d\Omega^{2}\;\;,\qquad
F(R)=(1-2M/R)\;\;,  \label{belso}
\end{equation}
where $R<2M$ and $T=bh$ are the time- and space-like coordinates,
respectively.

The canonical formalism of KS space-times is therefore equivalent to that of
the Schwarzschild solution. In the Hamiltonian theory of Schwarzschild black
holes we use a foliation consisting of spherical surfaces characterized by a
constant time parameter $t$, which is identified to the Schwarzschild time $
T $ [\citet{K}]. The geometry induced on these three-spheres has the form 
\begin{equation}
d\sigma^{2}=\Lambda^{2}(r)dr^{2}+R^{2}(r)d\Omega^{2}\;\;,
\end{equation}
where the functions $\Lambda$ and $R$ are chosen as canonical coordinates.
Then their conjugated momenta $P_{\Lambda}$ and $P_{R}$ are derived from the
action specified for the Schwarzschild space-time [\citet{K}] written in
terms of the canonical variables, 
\begin{equation}
P_{\Lambda}=-N^{-1}R(\dot{R}-R^{\prime}N^{r})\;\;,\qquad P_{R} =
-N^{-1}[\Lambda(\dot{R}-R^{\prime}N^{r})+R(\dot{\Lambda}-(\Lambda
N^{r})^{\prime})]\;\;.
\end{equation}
The Legendre transformation of the Lagrangian gives the Hamiltonian 
\begin{equation}
{\mathcal{H}}^{G}= P_{\Lambda}\dot{\Lambda}+P_{R}\dot{R}+NH^{G}+N^{r}H_{r}
^{G}  \label{HG}
\end{equation}
with super-Hamiltonian and supermomentum constraints 
\begin{equation}
H^{G}=R^{-1}P_{R}P_{\Lambda}+R^{-2}\Lambda P_{\Lambda}^{2}/2+\Lambda^{-1}RR^{\prime\prime} \\
-\Lambda^{-2}RR^{\prime}\Lambda^{\prime}-\Lambda^{-1}R^{\prime2}/2-\Lambda/2\;,
\end{equation}
\begin{equation} H_{r}^{G}  =P_{R}R^{\prime}-\Lambda P_{\Lambda}^{\prime}\;.
\end{equation}

The basics of the Hamiltonian formulation do not change if we couple matter
fields to gravity. We only have to enlarge the phase space of gravity by
including the canonical variables of the matter sources. After decomposing
the matter action we can derive the Hamiltonian for the matter fields as
well. Then the constraint equations of gravity must be supplemented with
those of the matter fields, which gives the full of set constraints imposed
on the total system.

\section{Kantowski-Sachs cosmologies with anisotropic fluid}

Exact solutions for KS space-times with anisotropic fluid sources have also
been found in the form 
\begin{equation}
ds^{2}=-2ae^{L^{2}}RdL^{2}+ae^{L^{2}}R^{-1}dZ^{2}+R^{2}d\Omega^{2}\;\;,\qquad a=-1\;\;,  \label{gmet}
\end{equation}
where 
\begin{equation}
-R=a(e^{L^{2}}-2L\Phi_{B})\;\;,\qquad\Phi_{B}=B+\int^{L}e^{x^{2}}dx\;\;,
\end{equation}
with $L$ and $Z$ the time and the radial coordinates $L$ and $Z$ [
\citet{homogen}]. The time dependence of the metric components shows that
the KS cosmology with anisotropic fluid is not static. By considering the
time evolution of the radial length $R(L)$ and the co-moving energy density
of the Universe, we see that the KS Universe has a finite lifetime with an
initial and a final singularity.

Anisotropic fluids can be considered as superpositions of two cross-flowing
null dust streams [\citet{L}]. Thus the action of an anisotropic fluid with
the co-moving density $\rho$, the four velocity $U^{\alpha}$ of the fluid
particles and a vector field $X^{\alpha}$ describing the direction of the
pressure forces is given by 
\begin{equation}
S^{F}=-\frac{1}{2}\int dx^{4} \sqrt{-g}\rho(U^{\alpha}U_{\alpha}+ X^{\alpha
}X_{\alpha})\;\;,  \label{enimp}
\end{equation}
which is both algebraically and dynamically equivalent to the action of the
two colliding null dust flows with the four velocities $u^{\alpha}$ and $
v^{\alpha}$: 
\begin{equation}
S^{2ND}=-\frac{1}{2}\int dx^{4}\sqrt{-g} \rho(u^{\alpha}u_{\alpha}+v^{\alpha
}v_{\alpha})\;\;.  \label{enimp2}
\end{equation}
Here the same energy density is chosen for both the dust components so that
the net flow should vanish for the static configuration.

A canonical formalism was previously developed for two cross-flowing null
dust streams coupled to the geometry by [\citet{BH}], but did not solve the
problem of the absence of a time standard for the colliding null dusts. The
anisotropic fluid interpretation of two in- and outgoing null dust streams
indicates there may be a possibility to use the same procedure as in the
case of the incoherent dust in order to find an internal time for the
canonical dynamics of colliding null dust streams.

The action with the constraint equations for the spherically symmetric
vacuum solution is to be supplemented with those of the matter fields. If we
write of the null vector fields in terms of the coordinates $Z$ and $L$, 
\begin{equation}
u_{\alpha}=WZ_{,\alpha}/\sqrt{2}+RWL_{,\alpha}\;\;,\qquad v_{\alpha
}=WZ_{,\alpha}/\sqrt{2}-RWL_{,\alpha}
\end{equation}
and make the same decomposition for the vector fields $U^{\alpha}$ and $
X^{\alpha}$, 
\begin{equation}
U_{\alpha}=WZ_{,\alpha}\;\;,\qquad X_{\alpha}=\sqrt{2}RWL_{,\alpha}
\end{equation}
with $W=(ae^{L^{2}}/R)^{1/2}$, the matter actions (\ref{enimp}) and (\ref
{enimp2}) can be expressed with these coordinates as well. By extremizing
these actions with respect to the variables $Z$, $L$ and the parameter $\rho$
, we obtain equivalent equations of motion for the null dust and fluid
models.

We use the coordinates $L(t,r)$ and $Z(t,r)$ as the canonical variables for
the matter and derive the canonical momenta conjugated to them form the
matter action: 
\begin{equation}
P_{L}=2a\sqrt{g}R^{2}\frac{\rho W^{2}}{N}(\dot{L}-N^{r}L^{\prime
})\;\;,\qquad P_{Z}=a\sqrt{g}\frac{\rho W^{2}}{N}(\dot{Z}-N^{r}Z^{\prime
})\;\;.
\end{equation}
As a result of the Legendre transformation of the Lagrangian in the matter
actions (\ref{enimp}) and (\ref{enimp2}), we obtain the same Hamiltonian for
both types of matter sources 
\begin{equation}
{\mathcal{H}}^{M}=P_{L}\dot{L}+P_{Z}\dot{Z}+NH^{M}+N^{r}H_{Z}^{M}\;\;,
\label{H2ND}
\end{equation}
where the super-Hamiltonian and supermomentum constraints imposed on the
matter variables are 
\begin{equation}
H_{\bot }^{M} =\left[ \frac{1}{2\sqrt{g}\rho W^{2}}\left( P_{Z}^{2}+\frac{1}{2aR^{2}}P_{L}^{2}\right) +\sqrt{g}\frac{\rho W^{2}}{2\Lambda ^{2}}\left((Z^{\prime })^{2}+2R^{2}(L^{\prime })^{2}\right) \right] \;\;, 
\end{equation}
\begin{equation}
H_{r}^{M} =Z^{\prime }P_{Z}+L^{\prime }P_{L}\;\;.
\end{equation}
The Hamiltonian (\ref{HG}) of the vacuum equations, together with Eq. (\ref
{H2ND}) describe the KS space-time with two colliding null dust streams or
equivalently, an anisotropic fluid source. The super-Hamiltonian and
supermomentum constraints of the total system are given by 
\begin{equation}
H_{\bot } :=H_{\bot }^{G}+H_{\bot }^{M}=0\;\;,  \label{Hbot}
\end{equation}
\begin{equation}
H_{r} :=H_{r}^{G}+H_{r}^{M}=0\;\;.  \label{Hr}
\end{equation}
These constraints are replaced with an equivalent set by solving the old
constraints with respect to $P_{Z}$ and $P_{L}$. Hence the momenta
associated with the matter can be separated from the other variables in the
constraint equations (\ref{Hbot})-(\ref{Hr}): 
\begin{equation}
H_{\uparrow } =P_{L}+h(r;\Lambda ,R,L,Z,P_{\Lambda },P_{R})\;\;, 
\end{equation}
\begin{equation}
H_{\uparrow Z} =P_{Z}+h_{Z}(r;\Lambda ,R,L,Z,P_{\Lambda },P_{R})\;\;,
\end{equation}
where 
\begin{equation}
h =\sqrt{2a}RL^{\prime ^{-1}}\left[ \Lambda \sqrt{G}\frac{dZ}{dL}-\sqrt{2}
aR^{-1}H_{r}^{G}P_{L}Z^{\prime }\right] \left[ \left( \frac{dZ}{dL}\right)
^{2}+2aR^{2}H_{r}^{G}\right] ^{-1}\;\;, 
\end{equation}
\begin{equation}h_{Z} =-\sqrt{2a}RL^{\prime ^{-1}}\left[ \Lambda \sqrt{G}-\left( \sqrt{2}
aR\right) ^{-1}hZ^{\prime }\right] \;,
\end{equation}
\begin{equation}G^{2}=(H^{G})^{2}-g_{ab}H_{a}^{G}H_{b}^{G}\;\;.
\end{equation}
Since the momenta $P_{L}$ and $P_{Z}$ are separated from the rest of the
canonical variables, the algebra of the new constraints has strongly
vanishing Poisson brackets and the Dirac algebra of the old constraints
turns to an Abelian algebra of the new ones [\cite{BK}]. The time variable
introduced here can be useful in the description of radiation atmospheres of
stars consisting of the in- and outgoing radiation streams. Our result might
also provide better prospects for the canonical quantization of KS
cosmologies with cross-flowing null dust streams.

\section{Conclusion}

We have studied the Hamiltonian formulation of KS cosmologies. In the case
of vacuum, the KS space-time is equivalent to the exterior Schwarzschild
solution and we can use the canonical theory developed for Schwarzschild
black holes. In static KS space-times with spherical symmetry, filled with
an anisotropic fluid, the matter source is equivalent to two cross-streaming
radiations. Thus the proper time of the dust particles used in the
Hamiltonian treatment of the fluid space-times could also be introduced as a
time variable in the canonical formalism of the colliding null dust streams.
We have derived a new set of constraints for the fluid or colliding null
dust variables as well, in which the canonical momenta of the matter are
separated from the rest of the variables. As a result, we have obtained an
Abelian constraint algebra. Our treatment could give new possibilities for
the discussion of quantum KS cosmologies with anisotropic fluid or colliding
null dust streams.

\begin{acknowledgement}
We should like to thank Dr. L. \'A. Gergely for several discussions of the matters treated here. This work was supported by OTKA grants no. T046939 and TS044665.
\end{acknowledgement}

\end{document}